\documentclass[aps,pre,amsmath, amssymb,superscriptaddress, twocolumn]{revtex4-2}
\usepackage{graphicx} 
\usepackage{xcolor}
\usepackage{subcaption}
\usepackage{amsmath,amssymb,mathtools}
\usepackage[english]{babel}
\usepackage{lipsum}
\usepackage{caption}
\usepackage[utf8]{inputenc}

\usepackage{hyperref}

\DeclareMathOperator{\FTBE}{FTBE}

\begin{document}

\title{Topological Mixing and Braiding Universality in Polar Active Matter}

\author{Wei Feng}
\affiliation{School of Physics, Shaanxi Key Laboratory for Theoretical Physics Frontiers, Fundamental Discipline Research Center for Quantum Science and Technology of Shaanxi Province, Northwest University, 710127, Xi’an, China}

\author{Tianyu Ren}
\affiliation{School of Physics, Shaanxi Key Laboratory for Theoretical Physics Frontiers, Fundamental Discipline Research Center for Quantum Science and Technology of Shaanxi Province, Northwest University, 710127, Xi’an, China}

\author{Zhihan Ye}
\affiliation{School of Physics, Shaanxi Key Laboratory for Theoretical Physics Frontiers, Fundamental Discipline Research Center for Quantum Science and Technology of Shaanxi Province, Northwest University, 710127, Xi’an, China}

\author{Jonas Berx}
\email[]{jonas.berx@nbi.ku.dk}
\affiliation{Niels Bohr Institute,
University of Copenhagen, 2100 Copenhagen, Denmark}

\author{Guangyin Jing}
\email[]{jing@nwu.edu.cn}
\affiliation{School of Physics, Shaanxi Key Laboratory for Theoretical Physics Frontiers, Fundamental Discipline Research Center for Quantum Science and Technology of Shaanxi Province, Northwest University, 710127, Xi’an, China}

\date{\today}

\begin{abstract}
Connecting the autonomous kinematics of active matter to its emergent macroscopic transport is constrained by the requirement for high-resolution Eulerian velocity fields. Here, we use confined bacterial suspensions as a model active fluid, mapping the sparse Lagrangian trajectories of fluorescent \textit{spy} cells into (2+1)-dimensional geometric braids that directly encode the spatiotemporal entanglement of the flow. We use the finite-time braiding exponent (FTBE) as a proxy to quantify the topological entropy and chaotic mixing of confined bacterial suspensions. In moderately confined \textit{wet} systems, we find that hydrodynamic coupling drives a structural transition from a dilute active gas to coherent vortices, and ultimately to active turbulence, revealing three distinct density-dependent regimes of topological mixing. Conversely, truncating the hydrodynamic screening length via extreme confinement drives the system toward a \textit{dry} active matter limit. In this limit, dense in-plane steric collisions suppress irreducible entanglement and substantially reduce the FTBE at high particle densities. By evaluating the topological complexity generated per encounter, we reveal a transition from a discrete geometric encounter regime to an areal escape mechanism. Finally, we establish a square-root scaling between the FTBE and the effective diffusivity, placing the self-sustained mixing of active fluids into the pathline braiding universality class.
\end{abstract}

\maketitle

\section{Introduction}

Self-propelled particles such as bacteria continuously inject energy into their surrounding fluid environment at the microscopic level, driving self-organization~\cite{marchetti2013hydrodynamics,elgeti2014physics}. In particular, dense bacterial suspensions represent paradigmatic examples of living fluids, where emergent collective motion and \textit{active turbulence} are qualitatively distinct from their inertial counterparts~\cite{alert2022active, aranson2022bacterial}. These chaotic active flows significantly enhance macroscopic and anomalous transport of nutrients and metabolic biomass while regulating biochemical signaling pathways and quorum sensing, thereby facilitating collective survival ~\cite{wu2000,sokolov2009enhanced,Leptos2009,kearns2010field,mukherjee2021anomalous}.

The self-generated flows by dense microswimmers advect the swimmers themselves, giving rise to coherent structures that can substantially alter the surrounding flow field and the transport of embedded matter. Unlike \textit{dry} models of active matter, however, bacterial suspensions operate in a \textit{wet} environment, where strong hydrodynamic interactions (HI) mediate these collective behaviors~\cite{alert2022active}. Flagellated bacteria act as autonomous \textit{living stirrers} that generate local flow fields~\cite{lauga2015bacterial}, spontaneously yielding mesoscale coherent flows through hydrodynamic coupling. Historically, the enhanced mixing in such active baths has primarily been quantified using macroscopic statistical approaches, including metrics such as mean squared displacements (MSD) and effective diffusion coefficients~\cite{wu2000,kurtuldu2011enhancement,Guasto2010,Mino2013}.
These phenomenological metrics quantify the macroscopic consequences of enhanced transport, while obscuring the geometric micromechanisms driving it~\cite{abbaspour2021enhanced, pushkin2013fluid}. In classical flows, nonlinear folding of fluid elements is constrained by spatial and temporal delays, typically emerging only after prolonged linear stretching~\cite{kelley2011separating}. Motile microswimmers can potentially bypass this bottleneck through deterministic spatial entanglements that rapidly deform material structures and enhance fluid dispersion. Yet, how these microscopic geometrical interactions give rise to anomalous transport at larger scales remains poorly understood. 

\begin{figure*}[t]
    \centering
    \includegraphics[width=0.85\textwidth]{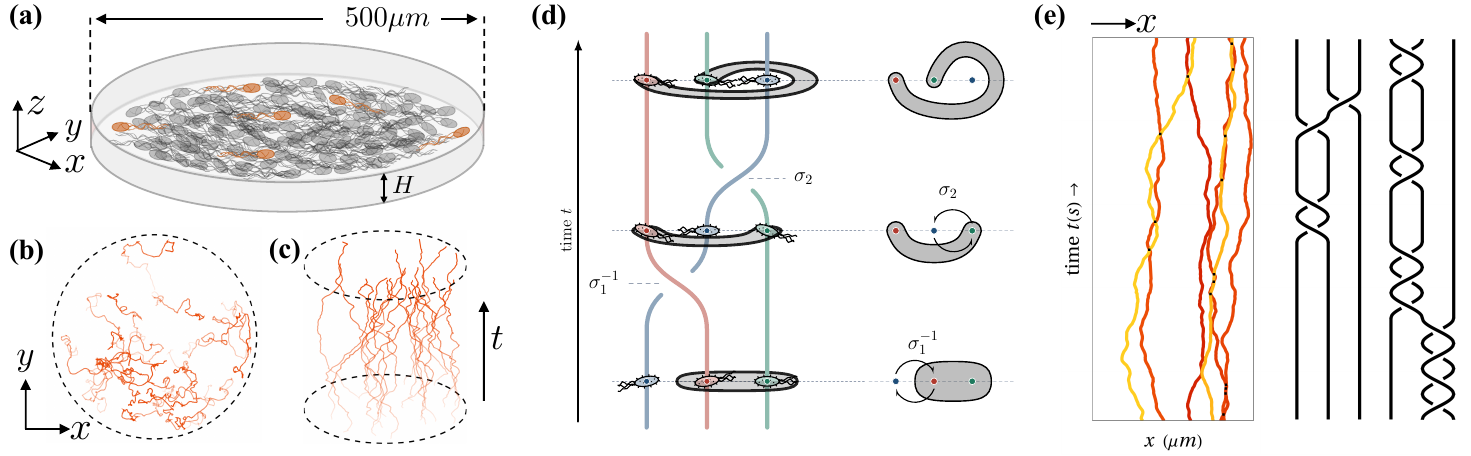}  
    \caption{\textbf{Self-propelled bacteria as living stirring particles in confinement.} 
    (\textbf{a-c}) Experimental setup and trajectory reconstruction of confined swimming bacteria. Fluorescently labeled \textit{spy} bacteria (orange) are tracked over time, with each trajectory represented by a corresponding orange line. (\textbf{d}) Sketch showing the braid crossing operators $\sigma_i^{\pm1}$ as well as the material line stretching (black) due to bacterial movement. (\textbf{e}) Projection of six typical \textit{spy} trajectories (colored lines) on the $x-t$ axis, showing the unsigned crossings (small black points), as well as the corresponding braid diagram. 
    }
    \label{fig:Setup_Sketch_Braids}
\end{figure*}

A powerful mathematical framework to study these geometric origins lies in topological fluid mechanics. By tracking individual Lagrangian trajectories and combining them into a mathematical braid, this approach quantifies the stretching and folding of material lines driven by stirring obstacles, establishing a rigorous mapping between discrete trajectory entanglements and the continuous topological deformation of the fluid via the Thurston-Nielsen theorem~\cite{boyland2000topological,thiffeault2006topology,thiffeault2010braids}. In particular, the finite-time braiding exponent (FTBE) provides a topological measure of the complexity generated by a flow over a finite observation time~\cite{Budisic2015}. Recently, the application of braid theory has opened new avenues in complex systems ~\cite{Allshouse2012,puckett2012trajectory,tumasz2013estimating,francois2015braid,smith2022topological,dilabbio2022braids,filippi2020using}, e.g., by characterizing winding statistics in \textit{dry} living flocks~\cite{2015PRLBraidingFlock,wang2025braided} or chaotic flows in active nematics driven by topological defects~\cite{tan2019topological,memarian2024controlling,mitchell2024maximally,firouznia2025self}. Bacterial active matter constitutes a distinct class of \textit{polar} active fluids, in which self-propelled swimmers act as living stirrers, generating active turbulence through the continual reorganization of mesoscale vortices and unsteady active circulations~\cite{dombrowski2004self,alert2022active,perez2026collective}. How Lagrangian trajectories entangle in such polar active turbulence, and whether their braid dynamics exhibit universal scaling across structural transitions, remains unknown. More broadly, it remains unclear whether a topological framework can capture the chaotic advection of \textit{wet} polar active turbulence, where strong HI and non-equilibrium energy injection govern the flow dynamics.

In this work, we use mathematical braid theory to characterize the Lagrangian topology of chaotic polar-active flows. A small fraction of fluorescently labeled \textit{spy} bacteria is doped into non-fluorescent host bacterial suspensions, serving simultaneously as energy-injecting stirrers and Lagrangian tracers. By constructing algebraic braids from sparse spy trajectories across a wide range of bacterial populations, we quantify the topological entropy of the active flow. We discover three topological mixing regimes in the wet active fluid and identify a topological transition from a dilute active gas to a dense, strongly correlated turbulent state. Ultimately, we establish a square-root scaling law between the topological entropy and the effective diffusivity. This provides experimental evidence that self-sustained polar active mixing falls within the \textit{Pathline Braiding Universality Class}~\cite{lester2024linking,Lester2025}, linking macroscopic superdiffusion to its microscopic topological origin.

\section{From trajectories to braids of self-stirrers}

At sufficiently high densities, the interplay between polar self-propulsion and hydrodynamic coupling drives bacterial suspensions into active turbulence~\cite{lauga2015bacterial, alert2022active}, generating chaotic flows that continuously stir the fluid. To experimentally capture this active stirring, we doped a sparse population of fluorescently labeled \textit{spy} bacteria (\textit{Escherichia coli} strain RP437, YFP-labeled) into a dense, non-fluorescent host population (\textit{E. coli} strain AW405) within quasi-two-dimensional micro-wells [Fig.~\ref{fig:Setup_Sketch_Braids}(a), Fig.~S1, videos~SM1-SM2 and~\emph{Supplementary Material}].

These \textit{spy bacteria} swim autonomously while simultaneously drifting in the active flow generated by the background host bacteria. The evolution of the trajectories of the spy bacteria reveals the chaotic stretching and folding of the complex flows of the active matter system. By tracking the \textit{spy} bacteria, we obtain trajectories
$\mathcal{Z}_n(t)=\{z_i(t)\}_{i=1,\dots,n}$, $t\in[0,T]$, with positions
$z_i(t)=(x_i(t),y_i(t))$ [Fig.~\ref{fig:Setup_Sketch_Braids}~(b), Fig.~S2], over an observation time $T$. Lifting time to a third axis turns each trajectory into a worldline [Fig.~\ref{fig:Setup_Sketch_Braids}~(c)]. 

The tracked trajectories form a \emph{geometric} braid in space-time. By isotopy, this braid can be represented as a standard \emph{mathematical} braid diagram by projection onto a plane [Fig.~\ref{fig:Setup_Sketch_Braids}(e)]. Crossings then generate a sequence of operators $\sigma_i^{\pm1}$, one per projected crossing [Fig.~\ref{fig:Setup_Sketch_Braids}(d)]. Kinematically, each operator $\sigma_i^{\epsilon_i}$ represents a single pairwise exchange event: two cells swap their in-plane order, with the sign $\epsilon_i$ recording which bacterium passes in front along the suppressed depth coordinate, i.e, $\sigma_i^{\pm1}$ for a clockwise ($\epsilon_i=-1$) or counterclockwise ($\epsilon_i=+1$) exchange of strands $i$ and $i+1$ [Fig.~\ref{fig:Setup_Sketch_Braids}(d)]. These operators generate the Artin braid group, and concatenating them gives the braid word $\beta_n=\sigma_{i_1}^{\epsilon_1}\cdots\sigma_{i_m}^{\epsilon_m}$, whose number of operators is the braid length $L=\sum_k|\epsilon_k|$: a direct count of pairwise crossings, and the simplest measure of stirring. Reducing the braid to its canonical (shortest) form removes trivial crossing events, lowering $L$ to the invariant $L_c$, which counts only irreducible entanglement.

\begin{figure*}[t]
    \centering
    \includegraphics[width=0.85\textwidth]{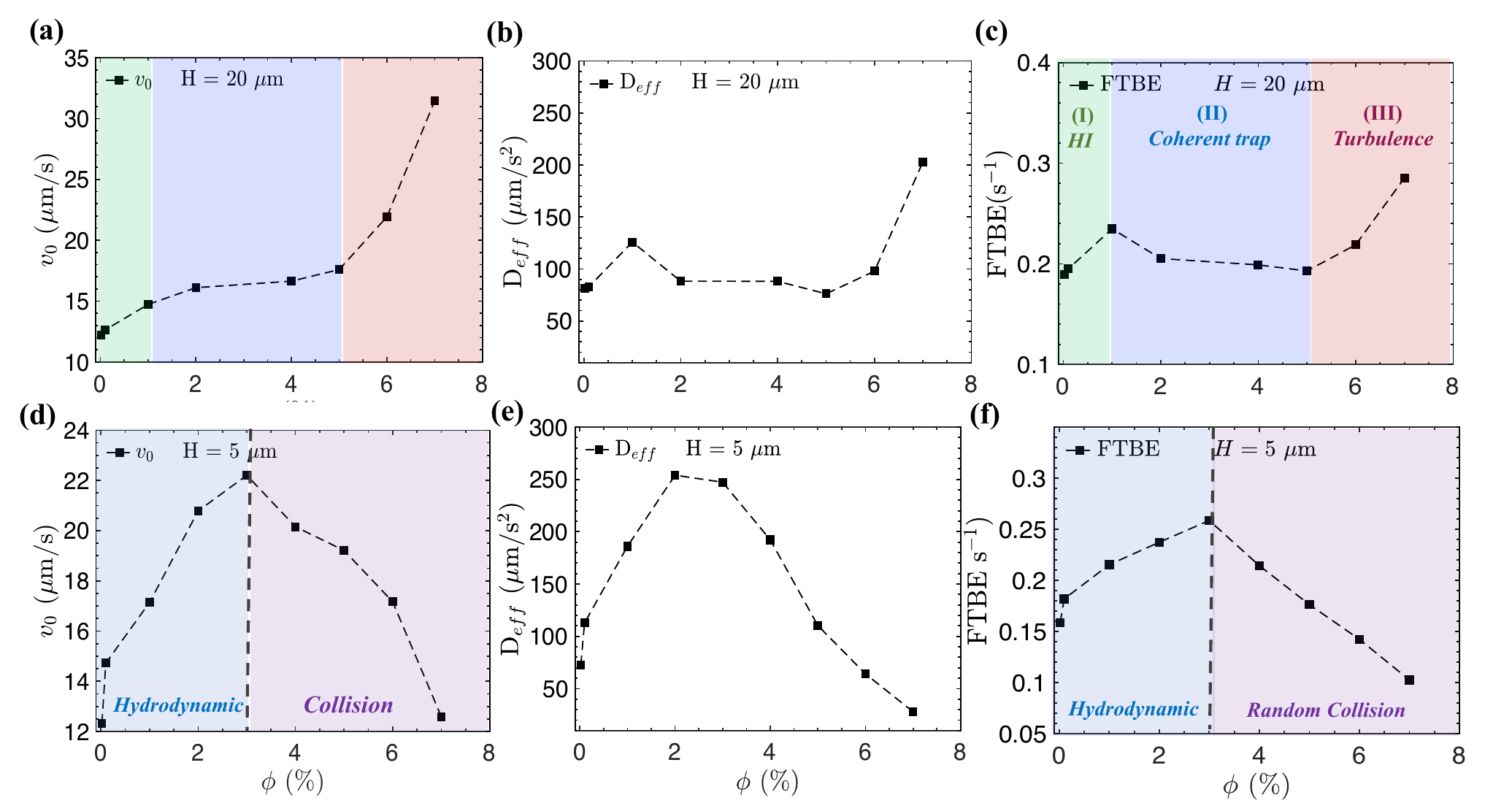}
    \caption{\textbf{Kinematic and braid-topological observables as a function of the volume fraction $\phi$.} {\bf(a--c)} Results for the $H=20\,\mu$m chamber: {\bf(a)} mean swimming speed $v_0$ of the cells; {\bf(b)} effective diffusion coefficient $D_{\rm eff}$ extracted from the post-ballistic regime of the MSD; and {\bf(c)} FTBE computed from braids constructed from $n=60$ \emph{spy} trajectories. {\bf(d--f)} Same observables as in (a--c), respectively, for the $H=5\,\mu$m chamber.}
    \label{fig:V_D_FTBE_2system}
\end{figure*}

The degree of mixing is quantified by the topological braid entropy $h_{\rm br}$, which measures the asymptotic growth rate of material-line length $\ell(t)$ of the fluid. A material filament lengthens as $\ell(t)\sim\ell_0\,e^{h_{\rm br}t}$. Note that $h_{\rm br}$ is a rigorous \emph{lower bound} on the topological entropy $h_{\rm top}$ of the underlying flow, since each cell is a moving obstacle that any comoving material line must wrap around. Positive $h_{\rm br}$ \emph{certifies} chaotic advection directly from the trajectories, with no velocity-field reconstruction required and independent of the near-field hydrodynamic and steric details of this \textit{wet} active matter. Instead of the asymptotic growth rate of the braid at $T,n\to\infty$, a measurable proxy is the finite-time braiding exponent (FTBE) for $n$ tracked \emph{spy} tracers~\cite{Budisic2015},

\begin{equation}
    \FTBE_n(\beta) = \frac{1}{T}\log\frac{|\beta \ell_E|}{|\ell_E|}\,,
    \label{eq:ComputeFTBE}
\end{equation}

\noindent where $\ell_E$ is a set of initial material loops encircling the cells at $t=0$, and the norm $|.|$ counts the number of intersections of the loops with the real axis~\cite{thiffeault2010braids,Budisic2015}. As the cells execute $\beta$ they drag these loops out in a manner similar to stretching rubber bands [see Fig.~\ref{fig:Setup_Sketch_Braids}(d)]. The FTBE is essential for mixing, with its finite-$n$ lower bound, $\FTBE_n\le h_{\rm top}$, becoming increasingly tight as $n$ and $T$ grow [Fig.~S10,~S11]. Its inverse, $\FTBE_n^{-1}$, therefore characterizes the timescale over which a bath of $n$ observed stirrers entangles the surrounding fluid [ref.~\emph{Supplementary Material}].

By varying the number density of bacteria and the degree of geometric confinement, we constructed geometric braids from tracked spy-bacteria trajectories across all experimental conditions [Fig.~S4, S5] and computed their FTBE with Eq.~\eqref{eq:ComputeFTBE}. Unlike apolar \textit{active nematics}~\cite{elgeti2014physics,tan2019topological}, the collective dynamics here are driven by the autonomous self-propulsion of \emph{polar} stirrers coupled with HI. By tightening the confinement, we can systematically screen this HI, effectively driving the system from a \textit{wet} active fluid to a \textit{dry} limit. This topological evaluation therefore enables us to directly unravel how the competition between polar motility, hydrodynamic coupling, and steric constraints dictates both the topological complexity and the physical transport properties of the flow.

\begin{figure*}[t]
     \centering
     \includegraphics[width=0.85\textwidth]{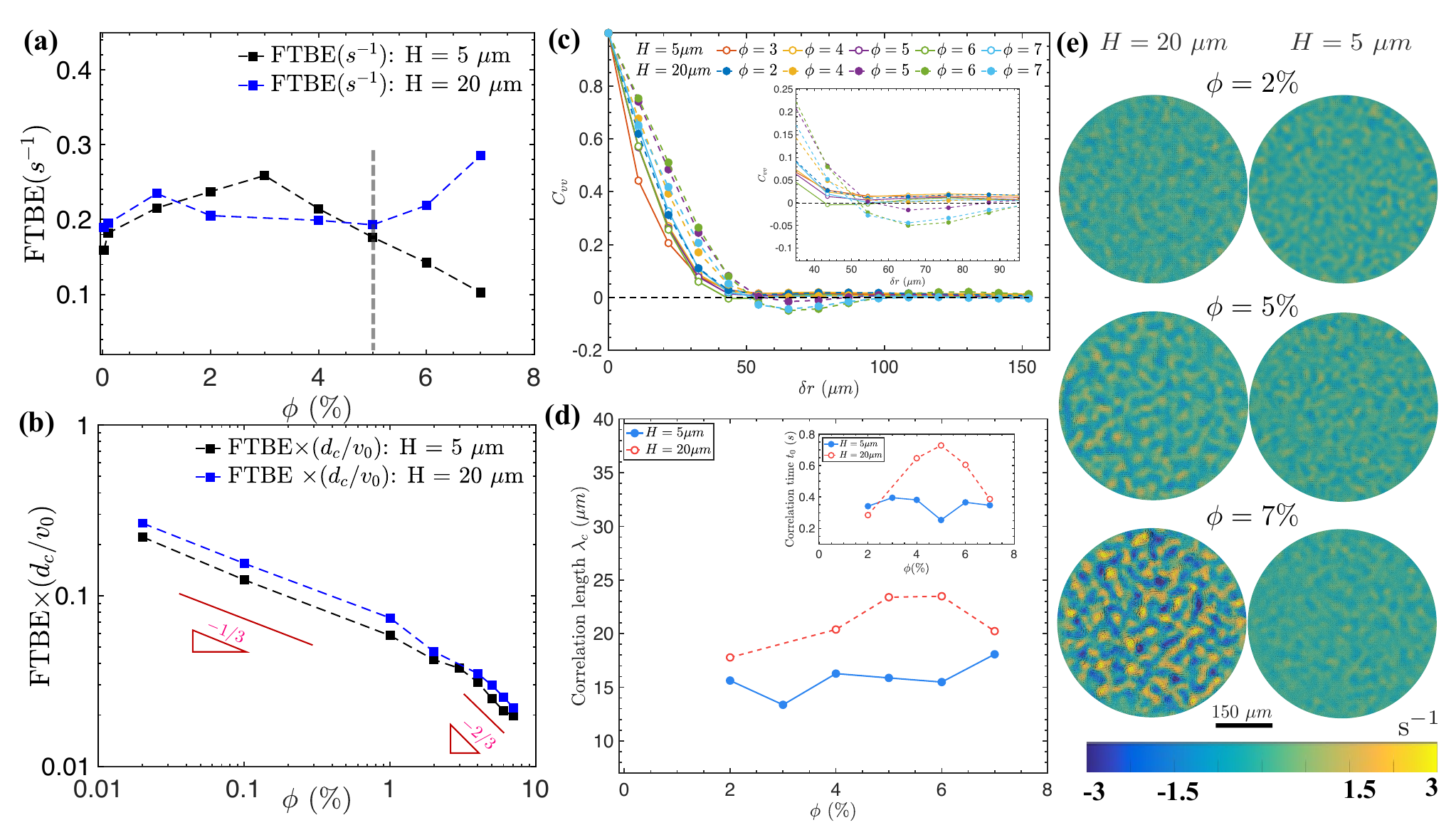}
     \caption{\textbf{Topological mixing and collective dynamics of the bacterial suspensions.}
{\bf(a)} Comparison of the FTBE between the $H=5\,\mu$m and $H=20\,\mu$m systems, with $n=60$ \emph{spy} tracers.
{\bf(b)} FTBE normalized by the characteristic traversal time $\tau_c=d_c/v_0$.
{\bf(c)} Velocity correlation function $C_{vv}$ of the active flow in the dense regime; the inset shows a magnified view of the negative minimum of $C_{vv}$.
{\bf(d)} Correlation length $\lambda_c$ and characteristic correlation time $t_0$ (inset, Fig.~S9) of the coherent flows. 
{\bf(e)} Velocity (vectors) and vorticity (colors) maps of the active flows at increasing volume fraction $\phi$ for both chamber heights. See also the Videos SM1-SM6.}
     \label{fig:Comparison}
 \end{figure*}
 
\section{Finite-time braiding from \textit{wet} to \textit{dry} active matter systems}\label{sec:braid_measures}

Hydrodynamic interactions between cells drive the emergence of coherent flows, competing with momentum screening imposed by the geometric confinement of the experimental chamber. Under strong confinement, momentum transfer is strongly screened by the nearby no-slip boundaries, suppressing hydrodynamic interactions beyond a length scale set by the confinement. Consequently, the balance between HI and confinement drives the suspension to undergo a crossover from a \textit{wet} to a \textit{dry} active system. For a \textit{wet} system of a quasi-two-dimensional circular well of height $H = 20\ \mu\text{m}$, we vary the host bacterial concentration $\phi$ and extract the swimming speed $v_0$ and effective diffusion coefficient $D_{\text{eff}}$ from the mean-squared displacement [Fig.~\ref{fig:V_D_FTBE_2system}(a,b); Fig.~S6]. The resulting dependence on $\phi$ exhibits three distinct regimes. In the dilute regime ($\phi \le 1\%$), both $v_0$ and $D_{\text{eff}}$ increase approximately linearly with concentration. This is consistent with weak HI that enhance advective transport, in agreement with previous simulations~\cite{hernandez2009dynamics,underhill2008diffusion,Leptos2009}. At intermediate concentrations, this growth slows and reaches a plateau, despite the suspension remaining continuously driven. Upon further increasing $\phi$, strong jets develop, and $v_0,\ D_{\text{eff}}$ rise sharply, marking the onset of chaotic active turbulence [Fig.~S7, Video SM1-SM3].

Conversely, under extreme confinement ($H = 5\ \mu\text{m}$), comparable to the size of a single bacterium, both $v_0$ and $D_{\text{eff}}$ increase rapidly in the dilute regime [Fig.~\ref{fig:V_D_FTBE_2system}(d,e); Fig.~S6] and reach a maximum near $\phi \approx 3\%$, corresponding to a mean inter-bacterial spacing of $d_c \sim 3\ \mu\text{m}$. Beyond this concentration, both quantities decrease monotonically. While Particle Image Velocimetry (PIV) reveals localized coherent flow structures near $\phi = 3\%$ [Fig.~S8, video~SM4-SM6], large-scale collective flows remain suppressed. At higher densities, even these localized structures progressively break down as frequent steric collisions disrupt collective motion. Consequently, with HI strongly screened, dynamics are dominated by in-plane steric collisions and boundary friction, yielding a continuous reduction in kinetic activity.

\begin{figure*}[t]
    \includegraphics[width = 0.85\textwidth]{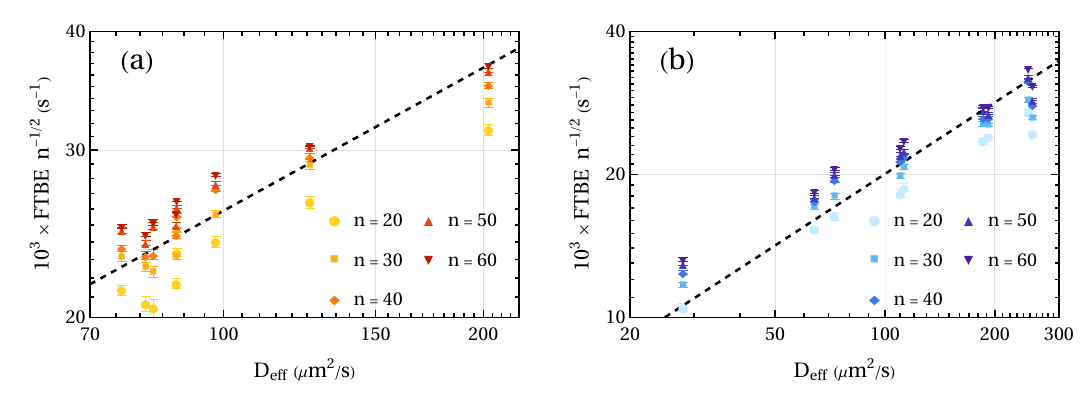}
    \caption{\textbf{Scaling of the FTBE with effective diffusivity $D_{\rm eff}$ for $n$ \emph{spy} tracers}, in {\bf (a)} the \emph{wet} ($H=20\,\mu\text{m}$) and {\bf (b)} \emph{dry} ($H=5\,\mu\text{m}$) experimental systems. Note that both are rescaled by $n^{-1/2}$ according to Eq.~\eqref{eq:sqrt} and presented on logarithmic scales, showing approximate data collapse consistent with the predicted scaling. Dashed black lines are guides for the eye with slope $1/2$.}
    \label{fig:Scaling}
\end{figure*}

The topological entanglement of the bacterial trajectories closely correlates with these macroscopic transport regimes. For the moderately confined \textit{wet} system ($H=20\ \mu\text{m}$), the FTBE displays three distinct $\phi$-dependent regimes [Fig.~\ref{fig:V_D_FTBE_2system}(c)]. In the dilute regime, the suspension behaves as an active gas. Trajectory crossing frequencies increase with concentration, and the FTBE rises in tandem with $v_0$. At intermediate concentrations, coherent flows develop [Fig.~\ref{fig:Comparison}(e)]. The characteristic scale of these structures is restricted by the vertical confinement, with the spatial correlation length $\lambda_c$ saturating near $20\ \mu\text{m}$ [Fig.~\ref{fig:Comparison}(d)]. In this regime, bacteria entrained by vortices orbit synchronously. Thus, their net contribution to the topological entropy remains minimal, causing the FTBE to saturate. These coherent structures destabilize and transition into fast active jets [Video SM1-SM3]. The proliferation of counter-rotating vortices—evidenced by the negative minimum in the velocity correlation function [Fig.~\ref{fig:Comparison}(c), inset]—alongside unsteady jet-like flows drives the suspension into active chaotic turbulence. The continuous breakdown and reorganization of these structures substantially enhance transverse crossings between particles in different fluid parcels, generating irreducible topological braids and driving an increase in the FTBE.

In contrast, under the extreme confinement of the \textit{dry} system ($H=5\ \mu\text{m}$), the topology-generating mechanisms are substantially altered [Fig.~\ref{fig:V_D_FTBE_2system}(f)]. The FTBE exhibits two distinct concentration-dependent regimes. For $\phi < 3\%$, advective entrainment enhances the swimming speed $v_0$ and produces an initial rise in the FTBE, akin to the dilute \textit{wet} system. At higher concentrations, however, crowding-induced collisions dominate, and the FTBE monotonically declines. This attenuation of topological entanglement reflects the strong hydrodynamic screening under extreme confinement: dense in-plane steric collisions continuously disrupt emerging localized flows, trapping bacterial trajectories in frequent rebounds and frustrated motions [Video SM4-SM6]. Topologically, these back-and-forth trajectories generate trivial crossings that algebraically cancel ($\sigma_i \sigma_i^{-1} = 1$), effectively suppressing the accumulation of topological entropy.

\section{Scaling and universality of Finite-Time Braid Entropy}\label{sec:scaling}

We define $\tau_c = d_c / v_0$ as the characteristic time required for a bacterium to traverse the mean inter-bacterial spacing. The dimensionless FTBE is then defined as
\begin{equation} 
    \text{FTBE}' = \text{FTBE} \times \left( \frac{d_c}{v_0} \right)\,.
\end{equation}
Physically, $\mathrm{FTBE}'$ quantifies the topological entropy generated over one characteristic encounter time. Experimentally, $\mathrm{FTBE}'$ exhibits two distinct scaling regimes as a function of the volume fraction $\phi$ [Fig.~\ref{fig:Comparison}(b)].

In the dilute limit, the suspension behaves as an ideal active gas, in which particles move nearly as independent ballistic random walkers. In this regime, the swimming speed depends linearly on $\phi$. Since the topological braid is constructed from a fixed number of spy bacteria, their geometric crossing rate is directly proportional to their swimming speed $v_0$. Thus, the ratio $\text{FTBE}/v_0$ remains constant, yielding $\text{FTBE}' \propto d_c$. Since $d_c \propto \phi^{-1/3}$, the dilute-limit scaling becomes $\text{FTBE}' \propto \phi^{-1/3}$, which agrees with a fitted exponent of -0.32 in the low-concentration regime [Fig.~\ref{fig:Comparison}(c)].

Conversely, at high concentrations labeled cells are transiently caged, either by coherent flow domains emerging from many-body hydrodynamic coupling or by steric neighbors under extreme confinement. In this regime bacteria may repeatedly recross without producing irreducible exchanges, increasing the raw crossing count $L$ while leaving the irreducible word length $L_c$ unchanged. Irreducible entanglement is produced only when a cell escapes across a domain boundary. Assuming diffusive relative motion with a typical displacement $\sim d_c$ per encounter, escaping a domain of size $\xi$ requires $(\xi/d_c)^2$ encounters. Thus, the probability $p$ of generating an irreducible exchange scales as $(d_c/\xi)^2$, yielding $\text{FTBE}' \propto p$.

Since $d_c \propto \phi^{-1/3}$ and the domain size $\xi$ is independent of $\phi$, this yields the high-concentration scaling $\mathrm{FTBE}' \propto d_c^{2} \propto \phi^{-2/3}$, which agrees well with the experimentally measured exponent of $-0.64$ [Fig.~\ref{fig:Comparison}(b)].
Scaled by the characteristic traversal time $\tau_c$, the dimensionless topological evolution reveals a crossover in the underlying entanglement dynamics of the active fluid: from a \textit{discrete geometric encounter regime} at low concentrations to a \textit{collective, correlated topological regime} at high concentrations. The $\phi^{-1/3}$ scaling reflects the dilute limit, where topological growth is governed by independent pairwise encounters and the efficiency of braid generation is determined by the mean separation between individual stirrers. In contrast, the $\phi^{-2/3}$ scaling arises in the dense regime, where coherent motion and steric constraints suppress independent exchanges, and irreducible entanglement is generated only through rare escape events from correlated domains. This crossover aligns with the emergence of collective motion in dense active bacterial suspensions reported previously~\cite{subramanian2009critical}.

The algebraic scaling between macroscopic transport and microscopic topology persists across both intermediate and extreme confinement. Despite the transition from weak hydrodynamic coupling to steric-collision dominance, the experimentally measured FTBE and $D_{\text{eff}}$ robustly follow the scaling law $\text{FTBE} \propto D_{\text{eff}}^{1/2}$ [Fig.~\ref{fig:Scaling}]. This universal scaling indicates that the active fluid belongs to the Pathline Braiding Universality Class~\cite{lester2024linking,Lester2025}.
This square-root scaling emerges analytically from a two-dimensional random-walk braiding model we briefly outline here; more details can be found in Ref.~\cite{lester2024linking}. Consider $n$ \emph{spy} trajectories within a domain of linear size $L_b$, taking statistically independent steps of magnitude $\Delta_T$ in random in-plane directions. Each step occurs over a time interval $\Delta t$, representing the mean time between braiding events. We define $\ell_b = L_b/\sqrt{n}$ as the mean spacing between observed trajectories.

For uncorrelated displacements in two dimensions, the mean-squared displacement is $\text{MSD}(t) = \Delta_T^2 t/\Delta t = 4 D_{\rm eff} t$, yielding $D_{\rm eff} = \Delta_T^2 /(4 \Delta t)$. While the FTBE converges to the Lyapunov exponent in the joint limit of long observation times and dense trajectory sampling~\cite{Budisic2015}, the rescaled FTBE of a random braid converges in the thermodynamic limit ($t\to\infty$, $n\to\infty$ at fixed $\ell_b$) to~\cite{lester2024linking}:

\begin{equation}
\frac{\ell_b \Delta t}{\Delta_T}\mathrm{FTBE}\longrightarrow\langle\lambda_{\sigma}\rangle
\simeq 0.8529,
\label{eq:lambdasigma}
\end{equation}

where $\langle\lambda_{\sigma}\rangle$ represents the topological entropy generated per braiding event by an infinite array of randomly braiding trajectories. Eliminating $\Delta_T$ using the expression for $D_{\rm eff}$ yields

\begin{equation}
  \mathrm{FTBE}
  =\frac{\langle\lambda_{\sigma}\rangle \Delta_T}{\ell_b \Delta t}
  =\frac{2\,\langle\lambda_{\sigma}\rangle}{L_b \sqrt{\Delta t}}n^{1/2}\,
   D_{\mathrm{eff}}^{1/2}.
  \label{eq:sqrt}
\end{equation}

The exponent $1/2$ reflects that particle dispersion accumulates as the square root of the number of braid events, whereas material stretching accumulates linearly. We note that the $n^{1/2}$ scaling saturates for large $n$: once the trajectory density resolves the velocity correlation length, the advective behavior is fully captured. Additional strands do not contribute further to the topological entropy, causing the FTBE to plateau~\cite{lester2024linking}. In the dilute limit, ballistic motion dominates ($\Delta_T\simeq v_0\Delta t$), reducing Eq.~\eqref{eq:sqrt} to $\mathrm{FTBE}=\langle\lambda_\sigma\rangle v_0/\ell_b$. This analytically recovers the constant ratio $\mathrm{FTBE}/v_0$ observed experimentally in Sec.~\ref{sec:braid_measures}.

This contrasts with electromagnetically driven 2D passive turbulence, where the braid entropy scales linearly with the single-particle dispersion coefficient ($\text{FTBE} \propto D_{\text{eff}}$)~\cite{francois2015braid}. That passive turbulence is multiscale and driven at a fixed energy-injection scale. Whereas, the single-scale, self-sustained active stirring considered here lacks such an externally imposed separation of scales. This difference in scaling exponents ($\propto D_{\text{eff}}^{1/2}$ versus $\propto D_{\text{eff}}$) might distinguish externally forced passive and self-generated active braiding.

\section{Conclusions}
In summary, we have introduced the mathematical framework of topological braiding into polar active fluids. By tracking a small fraction of \textit{spy} bacteria in a quasi-two-dimensional suspension and lifting their trajectories into a (2+1)-dimensional spatiotemporal manifold, we establish a direct mapping between the microscopic kinematics of autonomous living stirrers and the macroscopic topological entanglement of the active fluid. By systematically varying the bacterial concentration and vertical geometric confinement, we reveal how topological mixing is governed by the interplay between hydrodynamic interactions and steric constraints. In moderately confined \textit{wet} systems, enhanced HI at higher densities drives the emergence of large-scale coherent vortices and active jets. This collective motion promotes frequent transverse crossings between fluid parcels, driving a sharp increase in the FTBE. Conversely, extreme geometric confinement fundamentally reverses this topology-generating mechanism. By abruptly truncating the hydrodynamic screening length, extreme confinement suppresses macroscopic collective flows and kinetic activity. In this \textit{dry} limit, dense in-plane steric collisions and transient caging induce extensive algebraic cancellations in the braid group, effectively suppressing the FTBE.

We identify a topological transition from \textit{discrete geometric encounter} ($\phi^{-1/3}$) in dilute active gases to an \textit{areal escape} mechanism ($\phi^{-2/3}$) in dense, strongly correlated fluids. Furthermore, we establish a robust square-root scaling law relating the topological entropy and the effective diffusivity ($\text{FTBE} \propto D_{\text{eff}}^{1/2}$). This scaling collapses data across both wet and dry limits, placing the self-sustained mixing of active matter into the Pathline Braiding Universality Class and corroborating predictions from random braid theory. Our results offer insights into the transport and dispersion mechanisms governing complex living matter, and can inform the design of self-powered mixing and reconfigurable active metamaterials.

\begin{acknowledgments}
The authors acknowledge the funding support from the National Natural Science Foundation of China (No.12174306), the Natural Science Basic Research Program of Shaanxi (2023-JC-JQ-02), Shaanxi Academy of Fundamental Sciences (Mathematics, Physics No.23JSY024), the National Natural Science Foundation of China No. 12474197 and the Shaanxi Youth Science and Technology Star Project - 2025ZC-KJXX-51. J.B. thanks D. Lester for sharing a corrected version of Ref.~\cite{lester2024linking}.

\end{acknowledgments}

\bibliography{biblio}

\end{document}


\title{Supplementary material: Topological Mixing and Braiding Universality in Polar Active Matter}

\author{Wei Feng}
\affiliation{School of Physics, Shaanxi Key Laboratory for Theoretical Physics Frontiers, Fundamental Discipline Research Center for Quantum Science and Technology of Shaanxi Province, Northwest University, 710127, Xi’an, China}

\author{Tianyu Ren}
\affiliation{School of Physics, Shaanxi Key Laboratory for Theoretical Physics Frontiers, Fundamental Discipline Research Center for Quantum Science and Technology of Shaanxi Province, Northwest University, 710127, Xi’an, China}

\author{Zhihan Ye}
\affiliation{School of Physics, Shaanxi Key Laboratory for Theoretical Physics Frontiers, Fundamental Discipline Research Center for Quantum Science and Technology of Shaanxi Province, Northwest University, 710127, Xi’an, China}

\author{Jonas Berx}
\email[]{jonas.berx@nbi.ku.dk}
\affiliation{Niels Bohr Institute,
University of Copenhagen, 2100 Copenhagen, Denmark}

\author{Guangyin Jing}
\email[]{jing@nwu.edu.cn}
\affiliation{School of Physics, Shaanxi Key Laboratory for Theoretical Physics Frontiers, Fundamental Discipline Research Center for Quantum Science and Technology of Shaanxi Province, Northwest University, 710127, Xi’an, China}

\date{\today}

\maketitle

\tableofcontents

\section{Experimental Methods}\label{sec:exp_methods}

\subsection*{Bacterial suspension preparation}
 
\emph{Escherichia coli} (strain RP437 labeled with YFP and strain AW405) were revived from frozen glycerol stocks stored at a temperature of $-80~^\circ$C and inoculated into 10 mL of standard Luria--Bertani (LB) medium containing 1.0\% tryptone, 0.5\% yeast extract, and 1.0\% NaCl. Cultures were incubated overnight at 30~$^\circ$C with shaking at 200 rpm. An aliquot of the overnight culture was then diluted into fresh LB to an initial optical density OD$_{600} \approx 0.05$ and grown for 5 hours to the mid-exponential phase, i.e, optical density OD$_{600} \approx 0.8$ for \textit{E. Coli} strain of AW405.

\begin{figure}[H]
    \centering
    \includegraphics[width=0.95\linewidth]{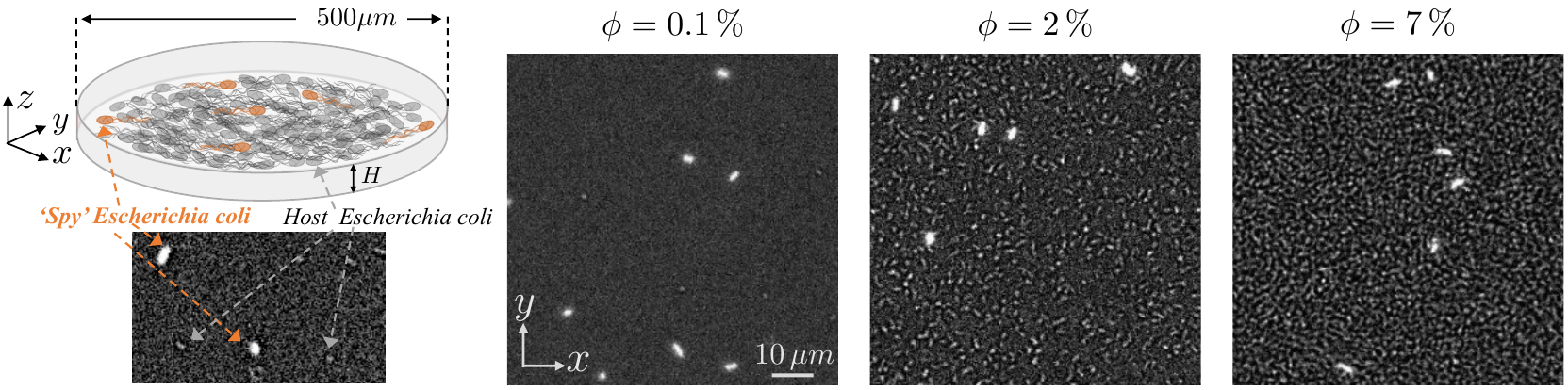} \caption{\textbf{Fluorescence snapshots of active bacterial baths at different volume fractions $\phi$.} Bright white dots represent fluorescently labeled \emph{spy} bacteria. Unlabeled host bacteria are faintly visible as gray structures due to residual background light scattering.}
    \label{fig:SiSetUp}
\end{figure}

Resuspended \textit{E. coli} strains RP437 and AW405 were mixed in motility buffer to obtain samples with final OD$_{600}$ values ranging from approximately 0.2 to 70, to obtain bacterial suspensions with varying volume fraction (density, or concentration). The OD$_{600}$ of \textit{E. coli} strain RP437 (the \emph{spy} bacteria) was kept constant at approximately 0.09 throughout the preparation, while the concentration of \textit{E. coli} strain AW405 (the host bacteria) was varied accordingly, as shown in Fig.~\ref{fig:SiSetUp}.

To determine the exact volume fraction of the bacterial suspensions with different OD, we need to specify the dimension of the individual bacteria. The typical bacterial body length was about $2~\mu$m and the diameter about $0.8~\mu$m. Note that OD is the quantity measured directly in the experiment, whereas its conversion to bacterial number density and subsequently to the volume fraction \(\phi\) is empirical. To quantitatively characterize the bacterial concentration, we establish a mapping between the measured optical density at 600 nm ($OD_{600}$) and the cellular volume fraction $\phi$. Based on standard empirical calibrations, an optical density of $OD_{600} = 1$ typically corresponds to a number density of approximately $10^9$ cells/mL ~\cite{kubitschek1990cell}. Assuming the bacterial body can be geometrically modeled with an average volume on the order of $1~\mu\text{m}^3$, this provides a reliable approximation where $OD_{600} = 1$ is equivalent to a volume fraction of approximately 0.1\%. Considering that we mainly focus on comparing conditions with substantially different OD values, we use a more convenient normalization for converting OD to \(\phi\). In our experiments, the bacterial suspension was carefully concentrated and diluted to systematically tune the volume fraction across a wide dynamic range, from a highly dilute state at 0.02\% to a densely packed, collective motion regime at 7\%. 

\subsection*{Experimental setup and microfluidic platform} \label{sec:setup}

Quasi-two-dimensional confinement was achieved using a microstructured polydimethylsiloxane (PDMS) layer placed on a glass substrate. The PDMS structures were fabricated by replica molding from a photolithographically patterned silicon master. Arrays of circular features with a diameter of $500~\mu\mathrm{m}$ and heights of either $20~\mu\mathrm{m}$ or $5~\mu\mathrm{m}$ were fabricated on silicon wafers using standard photolithography. The patterned silicon master was subsequently exposed to the vapor of trimethylchlorosilane to form an anti-adhesive, hydrophobic surface layer, thereby facilitating the release of the cured PDMS. Next, PDMS (Sylgard 184) was prepared by mixing the base and curing agent at a mass ratio of 8:1. The mixture was poured onto the silanized silicon master, whose perimeter was enclosed with tape to prevent the uncured PDMS from spreading. The PDMS precursor was degassed under vacuum for approximately $2~\mathrm{h}$ to remove trapped air bubbles and then cured at $80^{\circ}\mathrm{C}$ overnight. After curing, the PDMS layer was carefully peeled off from the silicon master, yielding a replica containing an array of circular wells with a diameter of $500~\mu\mathrm{m}$ and depths of either $20~\mu\mathrm{m}$ or $5~\mu\mathrm{m}$. The PDMS wells were subsequently treated with oxygen plasma to render their surfaces hydrophilic and facilitate the loading of the bacterial suspension. A $1.5~\mu\mathrm{L}$ aliquot of concentrated bacterial suspension was deposited onto a plasma-treated cover glass, and the microstructured PDMS layer was placed on top to form a sealed observation chamber, with the PDMS layer serving as the top wall and the glass cover slip as the bottom wall. The assembled glass--PDMS chamber was transferred to a humidity-controlled enclosure maintained at a relative humidity above $90\%$ [Fig.~\ref{fig:SiSetUp}] and subsequently mounted on the stage of an optical microscope for observation.

\begin{figure}[t]
    \centering
    \includegraphics[width=0.8\linewidth]{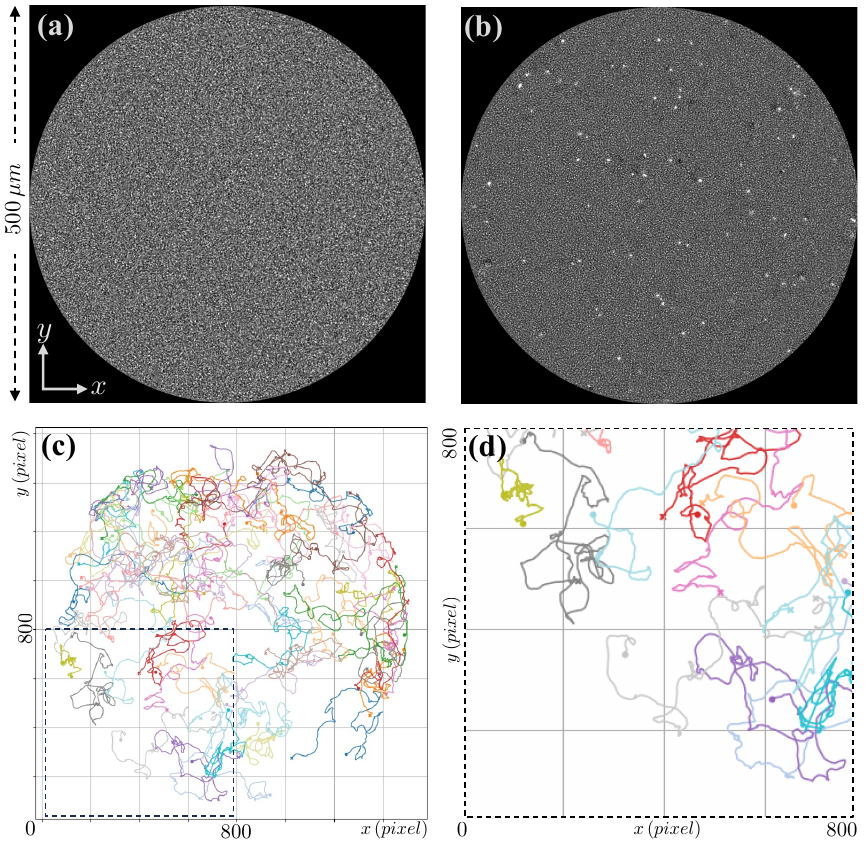}
    \caption{{\bf Tracking and connection of bacterial trajectories at a volume fraction of $\phi = 7\%$ in a chamber with height $H = 20\,\mu\text{m}$.} 
    (\textbf{a})~Bright-field image showing both host and spy bacteria. 
    (\textbf{b})~Fluorescence image of the same region, where the bright spots correspond to the fluorescent spy bacteria, leading to a high contrast between spy and host bacteria. 
    (\textbf{c})~Reconstructed trajectories of all identified spy bacteria. 
    (\textbf{d})~Magnified view of the boxed region in (c), where each line represents an individual spy bacterium trajectory (circles indicate starting points and `$\times$' markers denote endpoints).}
    \label{fig:longTrajectories}
\end{figure}

Bacterial suspensions were imaged using a Nikon Ti2-E inverted microscope equipped with a $20\times$ objective ($\text{NA} = 0.45$) and a high-speed camera (Hamamatsu ORCA-Flash4.0 V3, $2048\times2044$ pixels, $0.33\,\mu\text{m/pixel}$) under bright-field and fluorescence modes, as shown in Fig.~\ref{fig:longTrajectories}. 
To minimize the risk of misidentifying individual bacteria between consecutive frames, the imaging frame rate was set to $20\,\mathrm{fps}$. 

It is important to judiciously choose the frame rate for imaging, since we have to be careful to have enough time resolution to resolve the crossing events between swimming bacteria. The shortest time interval for a single crossing event is the perpendicular intersection of two bacteria, in such a way one bacteria swimming along its major axis is perpendicular to the transverse bacteria at its minor axis. Suppose that the bacteria has typical body length $L_b = 2\,\mu$m and diameter $d_b = 1\,\mu$m so that the distance of the bacteria running through is about $(L_b+d_b) \approx 3 \ \mu$m. Based on the maximum bacterial swimming speed measured in the experiments (Fig.~2a, $\sim 30\,\mu\mathrm{m/s}$), a bacterium would require approximately $0.1\,\mathrm{s}$ to travel one body length. Therefore, the chosen frame rate of 20 fps, i.e. 0.05 seconds per frame, is fast enough to distinguish each cross event.

\section{Particle Tracking Velocimetry at varying concentration}\label{sec:PTV}

\begin{figure}[t]
    \centering
    \includegraphics[width=0.8\textwidth]{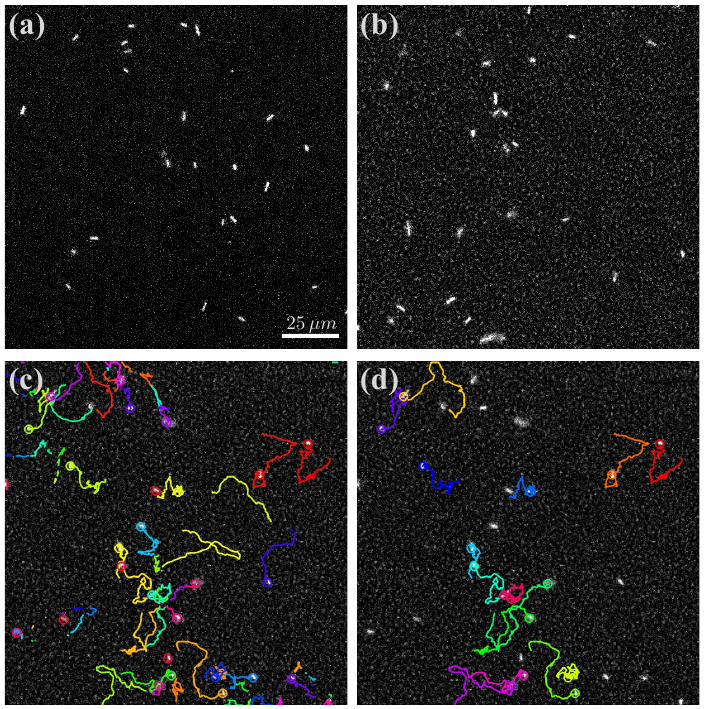}
    \caption{\textbf{PTV of spy bacteria in $H = 5\,\mu\text{m}$ chambers.} 
    (\textbf{a})~Fluorescence snapshot of spy bacteria at $\phi = 0.1\%$. 
    (\textbf{b})~Fluorescence snapshot of spy bacteria at $\phi = 7\%$. 
    (\textbf{c})~Trajectories of all spy bacteria from a 1$s$ movie shown in (\textbf{b}), where each line segment corresponds to a bacterial trajectory, and circles denote current bacterial positions. 
    (\textbf{d})~Reconnected and filtered long-time trajectories from (\textbf{c}), where only the spy bacteria satisfying the persistence threshold are shown.}
    \label{fig:reconnectTrajec}
\end{figure}

Raw spatiotemporal trajectories of fluorescent \emph{Escherichia coli} (strain RP437, YFP-labeled, with YFP excited by a 555\text{nm} laser) were extracted using the TrackMate plugin (v8.1.6)\cite{trackmate2017} in Fiji (v1.54p)\cite{fiji2012} with a nearest-neighbor linking algorithm, as shown in Fig.~\ref{fig:reconnectTrajec}. However, temporary defocusing and collisions frequently caused tracks to terminate prematurely and resume under new identifiers [Fig.~\ref{fig:reconnectTrajec}(c)].

Although the spy bacteria are dilute under all experimental conditions, they frequently undergo close spatial approaches, which poses significant challenges for continuous tracking following such events. While labeled spy bacteria can be readily tracked over short time intervals, obtaining sufficiently long trajectories is crucial for constructing the braids required in our FTBE calculation, see Eq.~(1) in the main text. The recording time window $T$ must substantially exceed both the duration of an individual near-pass event and the characteristic lifetime of coherent vortex structures in the active flow (typically $<1~\mathrm{s}$). Consequently, we choose $T > 10~\mathrm{s}$, corresponding to at least 200 consecutive frames per trajectory.

The algorithm sequentially addresses two primary modes of tracking failure:\\
(i) \textit{Near-pass resolution:} At each time step (frame), the spatial proximity between all particle pairs is computed. If the distance between two particles falls below a specified proximity threshold $ \approx 1 \mu m$, a near-pass event is inferred, and the corresponding trajectories are truncated from this exact frame to the frame when 2 particles separate far enough again to prevent cross-track switching.\\
(ii) \textit{Gap reconnection and velocity-based linkage:} To reconnect tracks fragmented by temporary defocusing or proximity truncation, an extensible search window is applied. For a track terminating at frame $t$, its instantaneous velocity vector $\mathbf{v}$ is estimated using up to $n$ preceding frames ($n \le 8$). The prospective spatial position at $t + \Delta t$ (where $\Delta t \le 2\text{ s}$) is linearly extrapolated as $\mathbf{r}_{\text{pred}} = \mathbf{r}(t) + \mathbf{v}\Delta t$. Fragmented segments initiating within a search radius $R_{\text{search}}$ ($R_{\text{search}} \le 16~\mu\text{m}$) centered at $\mathbf{r}_{\text{pred}}$ are identified as candidate links. A disjoint-set data structure (Union-Find) then evaluates these candidates to construct global, temporally non-overlapping trajectory groups. Finally, missing intermediate coordinates across temporal gaps are populated via linear interpolation to yield continuous trajectories.

To validate the algorithm, we artificially truncated trajectories generated by a machine-learning approach and compared against the ground-truth data. Under the benchmark testing conditions, the algorithm achieved a 100\% correction rate for cross-track switching and loss of tracking induced by bacterial collisions. For defocusing-induced tracking failures, the reconnection accuracy reached up to 89\%, depending on the defocus duration.

\begin{figure}[H]
    \centering
    \includegraphics[width=0.7\textwidth]{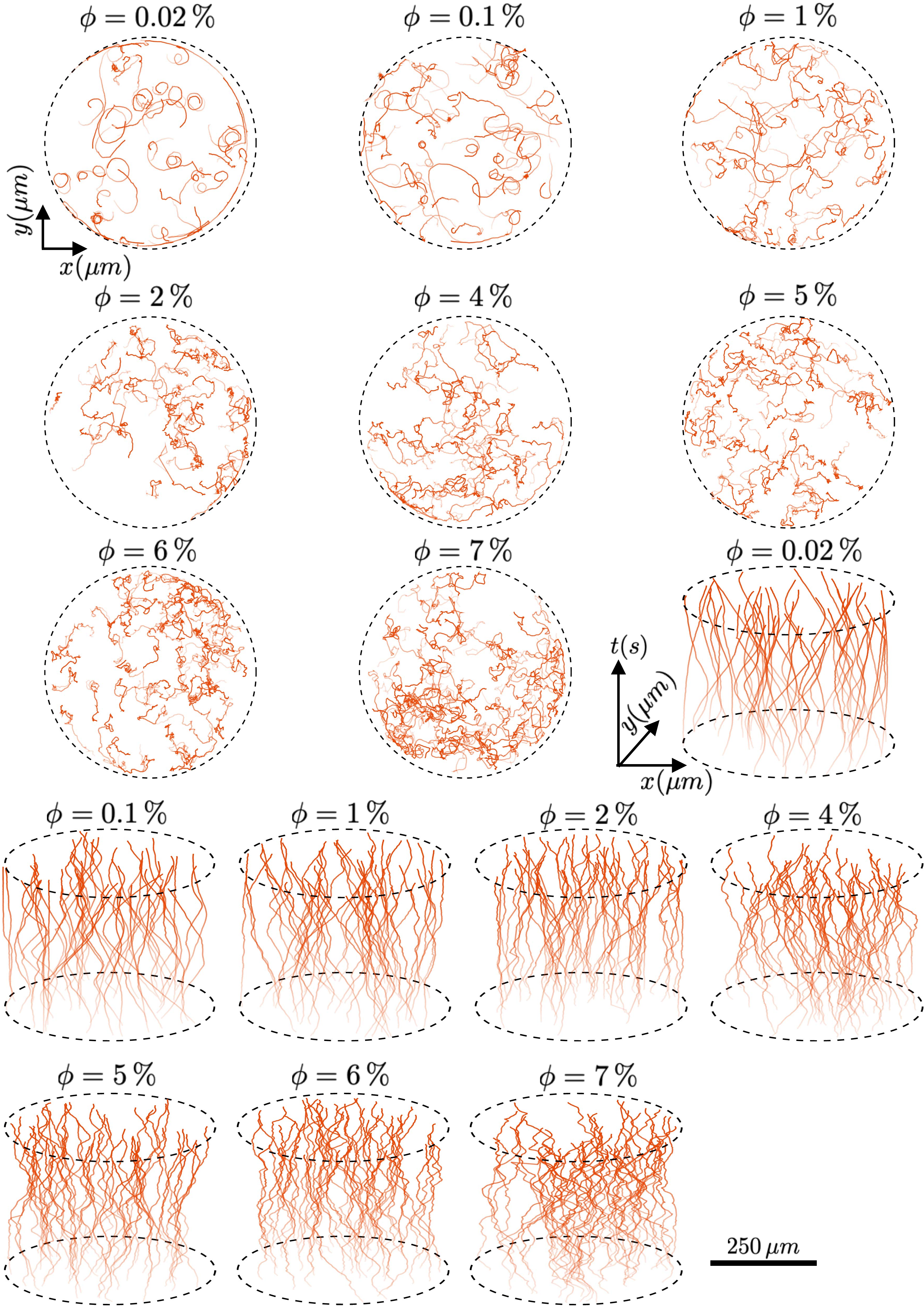}
    \caption{\textbf{2D ($x$--$y$) trajectories and corresponding space-time ($x$--$y$--$t$) braiding of spy bacteria at different $\phi$ in $H = 20\,\mu\text{m}$ chambers.} For each $\phi$, 60 continuous bacterial trajectories recorded within the same $15\,\text{s}$ time window are displayed, where each line represents the trajectory of an individual bacterium.}
    \label{fig:xy2xyt_H20}
\end{figure}

\begin{figure}[H]
    \centering
    \includegraphics[width=0.8\textwidth]{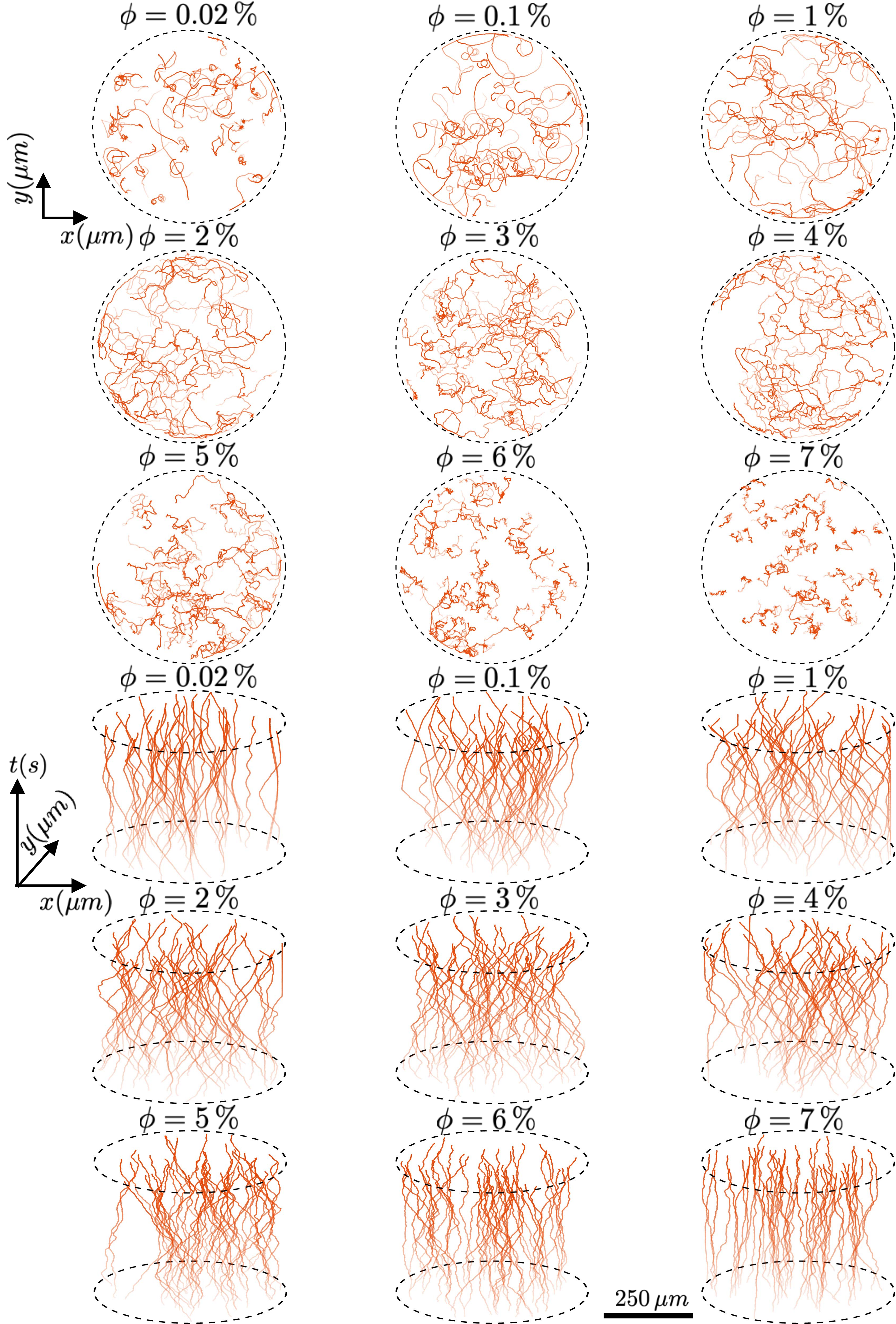}
    \caption{\textbf{2D ($x$--$y$) trajectories and corresponding space-time ($x$--$y$--$t$) braiding of spy bacteria at different $\phi$ in $H = 5\,\mu\text{m}$ chambers.} For each $\phi$, 60 continuous bacterial trajectories recorded within the same $15\,\text{s}$ time window are displayed, where each line represents the trajectory of an individual bacterium.}
    \label{fig:xy2xyt_H5}
\end{figure}

To illustrate the dynamical braiding states of bacteria under different bacterial concentrations in chambers with heights $H = 20\,\mu\text{m}$ and $5\,\mu\text{m}$, long trajectories of spy bacteria were extracted from the fluorescence images for analysis using particle tracking velocimetry (PTV), as shown in Figs.~\ref{fig:xy2xyt_H20} and~\ref{fig:xy2xyt_H5}. A set of 60 continuous trajectories of \emph{spy} bacteria were plotted within a fixed observation window of 15 s for each $\phi$. For $H = 20\,\mu\text{m}$, space-time ($x$-$y$-$t$) trajectory fluctuations progressively intensify with increasing $\phi$, qualitatively reflecting an enhanced degree of trajectory entanglement (braiding) [Fig.~\ref{fig:xy2xyt_H20}]. In contrast, for $H = 5\,\mu\text{m}$, the degree of trajectory entanglement exhibits a non-monotonic trend with increasing $\phi$, rising initially before declining at higher volume fractions [Fig.~\ref{fig:xy2xyt_H5}].

\begin{figure}[H]
    \centering
    \includegraphics[width=0.95\textwidth]{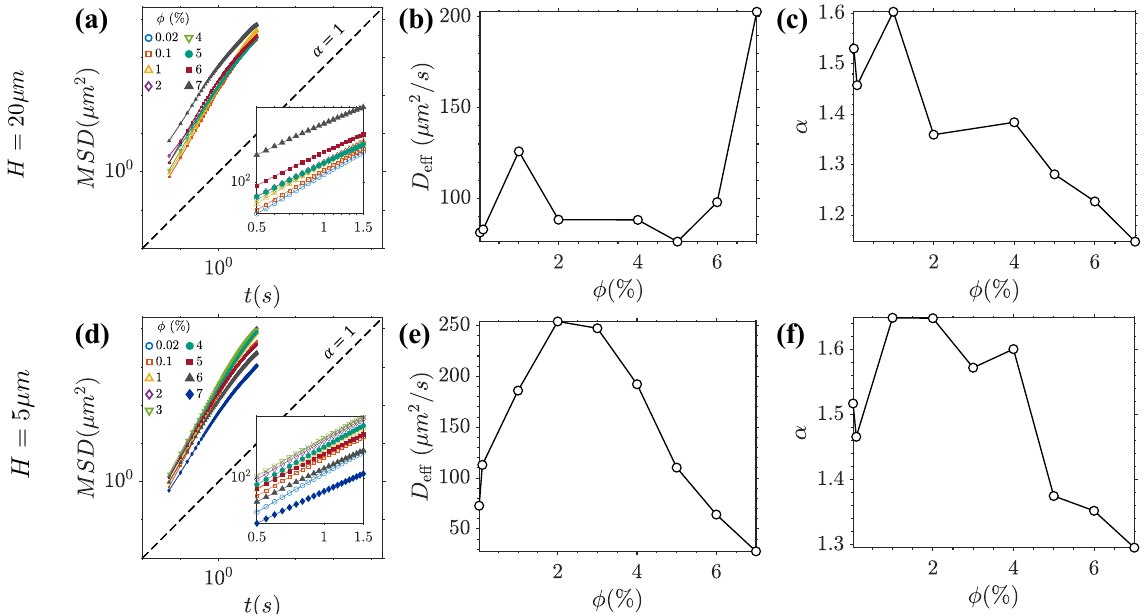}
    \caption{\textbf{Transport dynamics of spy bacteria across varying volume fractions $\phi$.} 
    (\textbf{a--c}) Chamber height $H = 20\,\mu\text{m}$: 
    (\textbf{a}) Mean-squared displacement (MSD) as a function of lag time $t$ on a log--log scale (inset shows a zoomed-in local region). 
    (\textbf{b}) Effective diffusion coefficient $D_{\text{eff}}$ as a function of $\phi$. 
    (\textbf{c}) Anomalous diffusion exponent $\alpha$ versus $\phi$. 
    (\textbf{d--f}) Corresponding results for height $H = 5\,\mu\text{m}$.}
    \label{fig:SI_diffusion}
\end{figure}

\section{Diffusion dynamics of spy bacteria}

To quantitatively evaluate the transport dynamics of bacteria across varying volume fractions ($\phi \in [0.02\%, 7\%]$) under chamber heights $H = 20\,\mu m$ and $5\,\mu m$, we analyzed the trajectories of individual \emph{spy} bacteria extracted from fluorescence image stacks. The results are shown in Fig.~\ref{fig:SI_diffusion}.

Single-bacterium trajectories, denoted as $z_i(t) = (x_i(t), y_i(t))$ for the $i$-th cell ($i = 1, 2, \dots, N$), are recorded at a temporal resolution of $\Delta t_{\text{frame}} = 0.05\,\text{s}$ (frame rate of 20 fps). The ensemble-averaged mean-squared displacement (MSD) as a function of lag time $\Delta t$ is defined as:
\begin{equation}
\text{MSD}(\Delta t) = \frac{1}{N} \sum_{i=1}^{N} \left\langle |z_i(t + \Delta t) - z_i(t)|^2 \right\rangle_t,
\end{equation}
where $\langle \cdot \rangle_t$ denotes time-averaging over time origins $t$. To ensure statistical robustness, trajectories with durations shorter than $4\,\text{s}$ are excluded. All trajectories are truncated at a maximum duration of $10\,\text{s}$, with at least 60 trajectories per volume fraction ($\phi$) spanning the full $10\,\text{s}$ duration.

To characterize transport beyond the initial ballistic regime, we analyze the MSD over the intermediate time interval $\Delta t \in [1.0, 4.0]\,\text{s}$. The effective two-dimensional diffusion coefficient $D_{\text{eff}}$ and the anomalous diffusion exponent $\alpha$ are extracted using:
\begin{align}
\text{MSD}(\Delta t) &= 4 D_{\text{eff}} \Delta t + C, \label{eq:Deff_fit} \\
\ln \left[ \text{MSD}(\Delta t) \right] &= \alpha \ln(\Delta t) + \ln K, \label{eq:alpha_fit}
\end{align}
where $C$ and $K$ are fitting parameters.
\section{Particle Image Velocimetry of active flows}\label{sec:PIV}

To characterize the collective bacterial motion at different bacterial concentrations, representative vorticity fields were obtained from the Particle Image Velocimetry (PIV) measurements for bacterial volume fractions ranging from $\phi=2\%$ to $7\%$. Bright-field image sequences are analyzed using PIV analysis for each volume fraction $\phi$, and 100 consecutive frames are imported at a frame rate of 20 frames per second (fps). Standard correlation robustness was selected for the correlation analysis. PIV analysis was performed using the PIVLabGUI app in MATLAB. For Pass 1, the interrogation area was 64 px with a step size of 32 px; for Pass 2, these values were 32 px and 16 px, respectively.

\begin{figure}[H]
    \centering
    \includegraphics[width=0.6\textwidth]{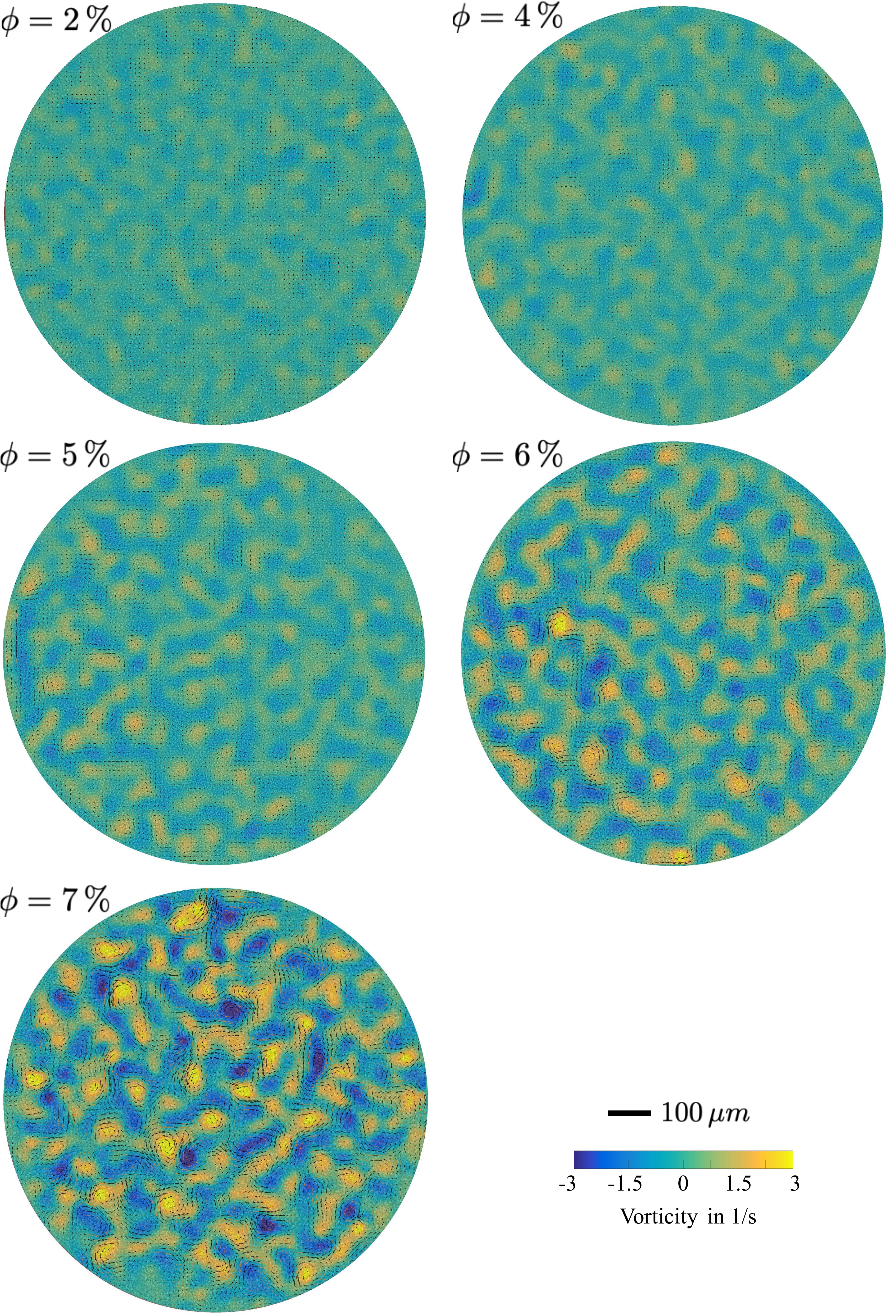}
    \caption{\textbf{PIV for the $H=20~\mu$m chambers at different bacterial volume fractions $\phi$.} The vorticity is indicated by a color scale while the velocity field is represented by small black and orange arrows.}
    \label{fig:all_PIV_H20}
\end{figure}

\begin{figure}[H]
    \centering
    \includegraphics[width=0.7\textwidth]{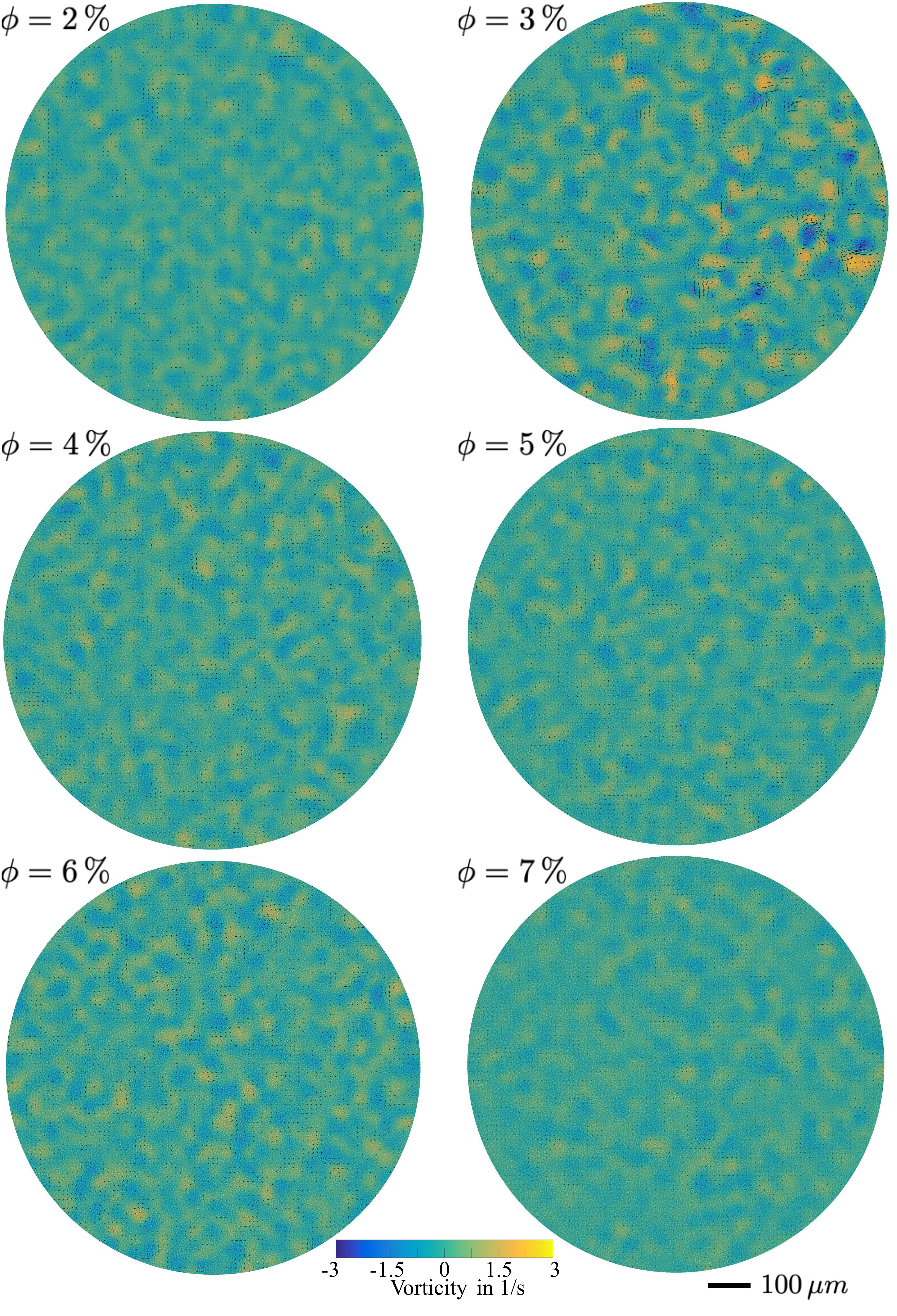}
    \caption{\textbf{PIV for the $H=5~\mu$m chambers at different bacterial volume fractions $\phi$.} The vorticity is indicated by a color scale while the velocity field is represented by small black and orange arrows.}
    \label{fig:all_PIV_H5}
\end{figure}

As shown in Fig~\ref{fig:all_PIV_H20} and Fig~\ref{fig:all_PIV_H5}, the vorticity fields exhibit spatially heterogeneous structures whose characteristic intensity and spatial organization vary with the bacterial volume fraction. In the $H=20~\mu$m chambers, at low $\phi$, the vorticity field is relatively weak and spatially diffuse, whereas increasing $\phi$ leads to increasingly pronounced spatial variations in vorticity. In particular, at higher bacterial volume fractions, coherent regions of positive and negative vorticity become more prominent, indicating enhanced collective flow fluctuations. For the $H=5~\mu$m chambers, the vorticity magnitude first increases and then decreases with increasing $\phi$.

\begin{figure}[H]
    \centering
    \includegraphics[width=0.6\textwidth]{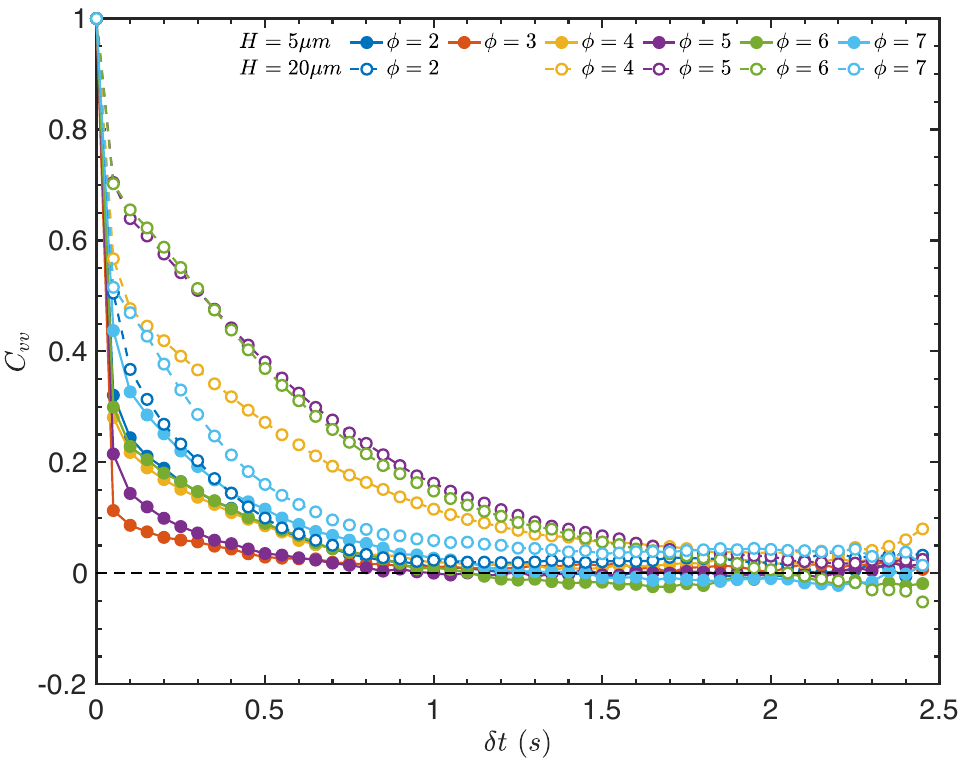}
    \caption{Velocity correlation function $C_{vv}$ of the active flow in the dense regime, derived from the PIV velocity fields. Filled (open) symbols indicate $H=5\,\mu$m ($H = 20\,\mu$m).}
    \label{fig:Cvv}
\end{figure}

To quantify the temporal persistence of the bacterial flow, we calculated the normalized temporal velocity autocorrelation function, $C_{vv}(\delta t)$, from the PIV-derived velocity fields. Here, $\delta t$ denotes the time lag between two velocity-field measurements. The resulting correlation functions for different bacterial volume fractions and chamber heights are shown in Fig.~\ref{fig:Cvv}.

\section{Braid Theory and Method}

An experiment $i$ at bacterial volume fraction $\phi_i$ yields a single set of particle trajectories, with each set possessing a different number of usable trajectories for the braid analysis. After the trajectory post-processing and reconnection algorithm explained in Section~\ref{sec:PTV} is applied, only a small fraction of trajectories contain data gaps or inconsistencies, which are filtered out. This leaves $N_i$ usable trajectories. These are subsequently analyzed using the {Braidlab} MATLAB library~\cite{Thiffeault2022-vk}.

In order for the braid measures to be consistent across experiments, a subset of $n$ trajectories must be chosen, with $n$ fixed. Since we typically have $N_i>n$ for all experiments $i$, we \emph{subsample} $M$ subsets of $n$ trajectories (a subbraid), in order to obtain sufficient statistics. In order to remove projection artifacts, each subbraid is projected along different projection angles $\theta\in[0,2\pi)$. For a single angle $\theta$, an ensemble average is first taken over subbraids. Subsequently, an angle average is taken to yield a single number for the braid measures, which can then be shown as functions of $\phi$.

For our data analysis, we set $n\leq 60$, $M=100$, and we project along 360 angles. The trajectories spanned a time interval of 15s, sampled at a rate of 20 frames per second. For each subbraid $\beta_n=\sigma_{i_1}^{\epsilon_1}\cdots\sigma_{i_m}^{\epsilon_m}$ consisting of $n$ strands, the braid length $L$, the writhe $W$ and the FTBE were computed using the following equations

\begin{align}
    L &= \sum_{k=1}^m|\epsilon_k|\,,\\
    W &= \sum_{k=1}^m \epsilon_k\,, \\
    \FTBE_n(\beta) &= \frac{1}{T}\log\frac{|\beta \ell_E|}{|\ell_E|}\,.\label{eq:ftbe}
\end{align}
Some comments on these measures are in order. To obtain the minimal braid length $L_c$ from $L$, a minimal braid representation is first determined from $\beta_n$ using Braidlab's \texttt{compact} function. Note that we also normalize the writhe by $n^2$, since it was shown~\cite{dilabbio2022braids} that it increases quadratically with increasing strand count. The writhe is not analyzed in detail in the main text, but serves as a sanity check for all data analysis; it is a measure for the \emph{global} rotation of the system and net handedness of the braid. Lastly, in Eq.~\eqref{eq:ftbe}, the norm $|.|$ counts the number of intersections between the loops $\ell_E$ and the real axis, corresponding to the intersections between the thick black material lines and the dashed line in Fig.~1(d) of the main text. As topological objects, the loops $\ell_E$ can be continuously deformed without consequence, provided no intersections are introduced, allowing them to be tightened around the bacteria they enclose. Their deformation under the braid can thus be viewed as that of rubber bands caught around the bacteria. The BraidLab library represents these loops using the integer Dynnikov coordinate system, described in Ref.~\cite{thiffeault2010braids}.

\begin{figure}[H]
    \centering
    \includegraphics[width=0.85\textwidth]{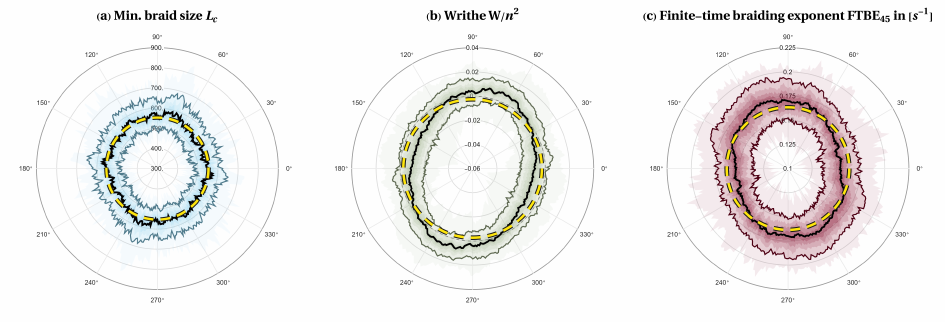}
    \caption{Braid-projection polar plots for the 20 $\mu$m chamber at volume fraction $\phi = 5\%$. Panels (a-c) display the minimal braid length $L_c$ (blue), rescaled writhe $W/n^2$ (green) and the $\FTBE$ (red), respectively, for braids constructed from $n=45$ particles. For each projection angle, an ensemble of 100 subsamples generates a distribution of values. Thick black curves show the ensemble averages, while thick colored curves mark the 10th and 90th percentiles. Dashed yellow lines indicate the values obtained after averaging over both subsamples and projection angles. Color opacity represents percentile levels, grouped into ten bands.}
    \label{fig:braidmeasures_H20phi5}
\end{figure}

In Fig.~\ref{fig:braidmeasures_H20phi5} we show a representative example of our full data at a volume fraction $\phi = 5\%$ for the $H=20\mu m$ system, with the averaging procedures indicated. The polar plots show the minimal braid size $L_c$ (blue), writhe $W/n^2$ (green) and FTBE (red) for different projection angles $\theta$. Per angle, the $M$ values are split into percentile ribbons, filled symmetrically about the median with opacity decreasing outward (i.e., densest at the median). The 10th and 90th percentile curves (colored curves), the subbraid-averaged curve (thick black), and the subbraid-angle-averaged curve (dashed yellow) are overlaid. It can be seen that the measures are quite robust with respect to the projection angle and subbraid sampling. For the writhe, Fig.~\ref{fig:braidmeasures_H20phi5} panel (b) shows that the mean is very close to zero, which is the expected null result for isotropic run-and-tumble motility in bulk 3D; this is not automatically guaranteed to vanish: individual \emph{E. coli} swim in clockwise circular trajectories near planar surfaces, because the force- and torque-free coupling between the counter-rotating body and flagellar bundle produces a clockwise-biasing torque under no-slip boundary conditions.

When a braid consists of periodic trajectories (i.e., the start- and endpoints are identified), changing the angle of projection does not change the braid entropy, since a rotation amounts to a \emph{conjugation} operation in the braid word $\beta$~\cite{Thiffeault2022-vk}. When the trajectories are not periodic, as in our experiments, the braid measures can change due to modified crossings at the start and end of the braid, depending on the projection angle. These fluctuations can be seen in Fig.~\ref{fig:braidmeasures_H20phi5} panels (b,c), and should become negligible for longer braids, again justifying our requirement for long, coexisting trajectories. It has been shown, in fact, that the relative deviation of angle dependence in the FTBE decreases approximately inversely with braid length~\cite{Budisic2015}. 

\begin{figure*}[htp]
    \includegraphics[width = 0.5\textwidth]{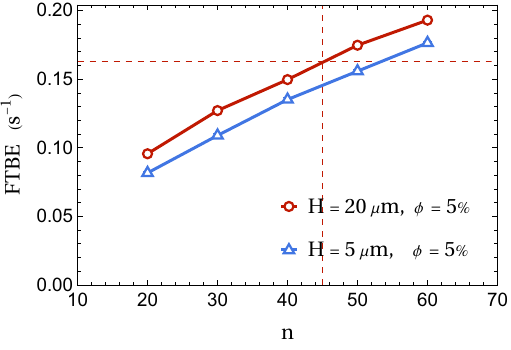}
    \caption{\textbf{Scaling of the FTBE with $n$ \emph{spy} tracers}, for the \emph{wet} ($H=20\,\mu\text{m}$, red) and \emph{dry} ($H=5\,\mu\text{m}$, blue) experimental systems. Both lines are monotonically increasing, as expected. Dashed red gridlines correspond to the average FTBE observed in Fig.~\ref{fig:braidmeasures_H20phi5}(c), with $n=45$.}
    \label{fig:ftbe_vs_n}
\end{figure*}

From the definition of the FTBE, Eq.\eqref{eq:ftbe}, it is also easy to see that it must be n non-decreasing function of $n$. Keeping the reference loop $\ell_E$ fixed, adding an extra stirrer can only increase (or at least not decrease) the intersection number; therefore it must hold that $|\beta_{n+1} \ell_E| \geq |\beta_n\ell_E|$. Since the logarithm is a monotonically increasing function, it follows that $\text{FTBE}_{n+1} \geq \text{FTBE}_n$. We show that this holds in our experiments as well in Fig.~\ref{fig:ftbe_vs_n}, for both chamber heights $H = 20\,\mu m$ and $5\,\mu m$ at density $\phi = 5\%$.

\section{Supplemental videos}

\textbf{Video SM1.} Bright-field microscopy video of bacteria in a
$H=20\,\mu\mathrm{m}$ chamber at $\phi=2\%$.

\textbf{Video SM2.} Bright-field microscopy video of bacteria in a
$H=20\,\mu\mathrm{m}$ chamber at $\phi=5\%$.

\textbf{Video SM3.} Bright-field microscopy video of bacteria in a
$H=20\,\mu\mathrm{m}$ chamber at $\phi=7\%$.

\textbf{Video SM4.} Bright-field microscopy video of bacteria in a
$H=5\,\mu\mathrm{m}$ chamber at $\phi=2\%$.

\textbf{Video SM5.} Bright-field microscopy video of bacteria in a
$H=5\,\mu\mathrm{m}$ chamber at $\phi=5\%$.

\textbf{Video SM6.} Bright-field microscopy video of bacteria in a
$H=5\,\mu\mathrm{m}$ chamber at $\phi=7\%$. \\

All videos have a spatial resolution of $0.33\,\mu\mathrm{m/pixel}$,
are recorded at $20\,\mathrm{fps}$, and contain 60 frames.